\documentclass[twocolumn, showpacs,showkeywords,preprintnumbers,amsmath,amssymb,aps]{revtex4}
\usepackage{bm}
\usepackage{latexsym}
\usepackage{eufrak}
\usepackage{ucs}
\usepackage[latin1]{inputenc}
\usepackage{latexsym}
\usepackage{amsmath}
\usepackage{epsfig}
\usepackage{ifpdf}
\usepackage{graphicx}
\usepackage{dcolumn}
\usepackage{amsmath}
\usepackage{amsfonts}
\usepackage{amssymb}
\usepackage{color}

\begin{document}

\title{Gravitational Waves Bounds in Brane-Worlds.}

\author{Miguel A. Garc\'{\i}a-Aspeitia} 
\email{aspeitia@fisica.ugto.mx}
\affiliation{Departamento de F\'isica, DCI, Campus Le\'on, Universidad de Guanajuato, C.P. 37150, Le\'on, Guanajuato, M\'exico}

\begin{abstract}
This paper is dedicated to investigate an astrophysical method to 
obtain the new dynamics generated by extra dimensions as well as bounds for the brane tension. Using the modified Einstein equations in the brane with a vanishing non-local effects, we study the contributions of the modified radiated power by gravitational waves and the stellar period modified by branes in a binary system composed by two neutron stars. Finally we propose two lower energy bounds, using these astrophysical methods.
 
\smallskip
\noindent \textmd{Keywords: Branes, Extra dimensions.}
\end{abstract}

\keywords{}
\draft
\pacs{04.50.-h}
\date{\today}
\maketitle

\section{Introduction.}
Brane world dynamics is an interesting alternative for complementing the Einsteinian general relativity (GR) by adding new degrees of freedom caused by the existence of extra dimensions. Many proposals try to solve diverse physical problems such as hierarchy problem \cite{Randall-I,Randall-II}, the cosmological constant (CC), the accelerated expansion of the universe \cite{DGP,Lue,Binetruy,Wang,Gogberashvili:1998vx} and the radiative corrections in quantum processes with an extra dimensions \cite{ADD,PerezLorenzana:2004na}. Similarly, other authors generalize, for example, the Randall-Sundrum models (RS) assuming a cosmological scenario with the aim of studying the space-time evolution, the scalar perturbations and the structure formation with brane world corrections \cite{aspeitia1,aspeitia2,sasaki,PhysRevD.62.024011} and possible consequences of a scalar-field as dark matter which only lives in a hidden brane \cite{aspeitia1,aspeitia2,sasaki}. 
Analogously, the perspective of interacting branes is welcomed due to the extra terms in the dynamical equations and in the equation of state (EoS) \cite{JuanLuis,Langlois}, which provide with relevant information principally in high energy epochs such as inflation \cite{JuanLuis}.

Numerous experimental and observational evidences constrain the range of action of the brane world dynamics \cite{sasaki,Roy, Germani,Rahaman,barrow,Ichiki,Bratt,Anupam}. For example, some observational results argue that the brane tension must be $\lambda>(1MeV)^{4}$ in order to nucleosynthesis holds \cite{Roy}; similarly, the dark radiation contributions caused by the non-local effects are constrained by CMB and nucleosynthesis observations to be no more than $\sim5\%$ of the radiation energy density \cite{Roy}. Other authors \cite{Germani,Rahaman}, focus in understanding the effects of extra dimensions and their repercussions on the astrophysical dynamics, principally on the modification of the Tolman-Oppenheimer-Volkoff (TOV) \cite{Germani} equation and in consequence in the galactic rotational curves and density profiles \cite{Rahaman}.

Particularly, this paper is focused on the assumption that the 5dim Einstein's equation is correct; the assumption generates corrections in the four dimensional Einstein's equation on the brane, producing quadratic contributions in the energy-momentum tensor and non local effects related with the Weyl's tensor and the fields contained in the bulk\cite{Shiromizu}. Remarkably these corrections must has effects in the CMB \cite{Koyama:2003be,Rhodes:2003ev}, gravitational waves (cosmic and astrophysical) \cite{Frolov:2002qm,Wands,Gorbunov:2001ge} and in galaxy dynamics imprints which could be measured in the future with space and terrestrial experiments \cite{LIGO, new}.

Under this scenario, we take the task of finding bounds to the brane tension to investigate the region of validity of the theory. We calculate the correction predicted by the brane theory about the power of gravitational waves radiated by an astrophysical binary system, and compare it with the observational data of the PSRB$1913+16$ \cite{weisberg} binary system, in order to set an upper bound of the brane tension. Also, it is calculated the equation of period and shown the new term provided by branes in high energy regime. Similarly, using the data of the binary system, it is possible to obtain an upper bound for the brane tension.

The article is organized as follows: In section \ref{Form}, we focus our attention to a short review of the mathematical branes formalism, in section \ref{GW} we obtain a bound of brane tension from an expression of the radiated power and orbital period, using an extensively studied binary system. Finally, in section \ref{CR} we discuss the results and give our conclusions.

\section{A short review of the Mathematical Formalism.}\label{Form}

Let us start by writing the equations of motion for an embedded brane
in a five dimensional bulk using the Randall-Sundrum II (RSII)
model \cite{Randall-II}, see also \cite{Roy,Shiromizu}. We first assume
that the Einstein equations are the gravitational equations of motion
of the 5dim Universe,
\begin{equation}
  G_{AB} + \Lambda_{(5)} g_{AB} = \kappa^{2}_{(5)} T_{AB} \, ,
\end{equation}
where $G_{AB}$ denotes the five dimensional Einstein tensor, $T_{AB}$
refers to the 5D energy-momentum tensor, $\Lambda_{(5)}$ represents
the 5dim cosmological constant, and $\kappa^{2}_{(5)}$ is the 5dim
gravitational coupling. In order to write the gravitational equations
of motion in the 4dim brane, we need to calculate the Gauss and
Codacci expressions, respectively,
\begin{subequations}
\begin{eqnarray}
  {}^{(4)}R_{\beta\gamma\delta}^{\alpha} =
  {}^{(5)}R^{\mu}_{\nu\rho\sigma} q_{\mu}^{\alpha}
  q_{\beta}^{\nu}q_{\gamma}^{\rho} q_{\delta}^{\sigma} +
  K_{\gamma}^{\alpha} K_{\beta\delta} - K^{\alpha}_{\delta}
  K_{\delta\gamma} \, , \\
  D_{\nu} K^{\nu}_{\mu} - D_{\mu}K = {}^{(5)}R_{\rho\sigma} n^{\sigma}
  q_{\mu}^{\rho} \, ,
\end{eqnarray}
\end{subequations}
where the extrinsic curvature over the 4D manifold $\mathcal{M}$ is
given by $K_{\mu\nu} = q_{\mu}^{\alpha} q_{\nu}^{\beta}
\nabla_{\alpha} n_{\beta}$, $K = K^{\mu}_{\mu}$, and $D_{\mu}$ is the
covariant differentiation of $q_{\mu\nu}$. It is important to remark that in the brane world scenario, our 4dim world is described by a domain wall (3-brane) ($\mathcal{M}$,$g_{\mu\nu}$) in 5dim space-time ($\mathcal{V}$,$q_{\mu\nu}$). We denote the vector unit normal to $\mathcal{M}$ by $n^{\alpha}$ and the induced metric on $\mathcal{M}$ by $g_{\mu\nu} = q_{\mu\nu}-n_{\mu} n_{\nu}$ \cite{Shiromizu}. Following an appropriate computation, it is possible to demonstrate that the modified 4dim Einstein's equation can be written as \cite{Roy,Shiromizu}
\begin{equation}
  G_{\mu\nu} + \xi_{\mu\nu} + \Lambda_{(4)}g_{\mu\nu} =
  \kappa^{2}_{(4)} T_{\mu\nu} + \kappa^{4}_{(5)} \Pi_{\mu\nu} +
  \kappa^{2}_{(5)} F_{\mu\nu} \, , \label{Eins}
\end{equation}
where
\begin{subequations}
\label{eq:1}
\begin{eqnarray}
  \Lambda_{(4)} &=& \frac{1}{2} \Lambda_{(5)} +
  \frac{\kappa^{4}_{(5)}}{12} \lambda^{2} \, , \label{lambda} \\
  \kappa^{2}_{(4)} &=& 8\pi G_{N} = \frac{\kappa^{4}_{(5)}}{6}
  \lambda \, , \label{eq:1b} \\
  \Pi_{\mu\nu} &=& - \frac{1}{4} T_{\mu\alpha}T^{\alpha}_{\nu} +
  \frac{TT_{\mu\nu}}{12} + \frac{g_{\mu\nu}}{24}
  (3T_{\alpha\beta}T^{\alpha\beta} - T^{2}) \, , \label{eq:1c} \\
  F_{\mu\nu} &=& \frac{2T_{AB} g_{\mu}^{A} g_{\nu}^{B}}{3} +
  \frac{2g_{\mu\nu}}{3} \left[T_{AB} n^{A}n^{B} - \frac{^{(5)}T}{4} 
  \right] \, , \label{eq:1d} \\
  \xi_{\mu\nu} &\equiv& {}^{(5)} C_{AFB}^{E} n_{E} n^{F} g_{\mu}^{A}
  g_{\nu}^{B} \, . \label{eq:1e}
\end{eqnarray}
\end{subequations}
Here, $\Lambda_{(5)}$ and $\Lambda_{(4)}$ are the five and four
dimensional cosmological constants, respectively, $G_{N}$ is
Newton's gravitational constant, $\kappa_{(4)}$ is the 4dim
gravitational constant, $\lambda$ is related to the brane tension,
$n_{E}$ is an unit normal vector to the brane manifold $\mathcal{M}$,
and ${}^{(5)}C_{AFB}^{E}$ is related with the 5dim Weyl's tensor \cite{Roy,Shiromizu}. Note that in what follows, we use the most-plus signature (${\rm diag}(-,+,+,+)$) for the
line elements, and natural units in which $c=\hbar=1$.
 
For purposes of simplicity, we will not consider bulk matter, which
translates into $F_{\mu\nu}=0$, and neglect the non-local effects 
caused by the Weyl's tensor, so that $\xi_{\mu\nu}=0$; the last hypothesis is due 
that the system of equations for the exterior is not closed until a further condition is given 
on the Weyl's tensor \cite{Germani,Roy}; we will comment on the effects this assumption has on the results. Note also that we will keep the presence of $\Pi_{\mu\nu}$, see Eq.~\eqref{eq:1c}, in the equations of motion throughout the paper, and these quadratic corrections will be our main concern for the study of brane effects on the 4D physical phenomena.

\section{Gravitational Waves Bounds.} \label{GW}

In this section, we study the gravitational waves originated in binary systems in the brane world context, using the modified Einstein's Eq. \eqref{Eins}. At first approximation, we assume that contributions of the non-local effects caused by the Weyl tensor reflected in $\xi_{\mu\nu}$ can be neglected \footnote{In general $\xi_{\mu\nu}\neq0$ in realistic astrophysical models, however it is necessary to note that the system of equations for the exterior is not closed until a further condition is given on the Weyl tensor; for this reason $\xi_{\mu\nu}=0$ \citep{Roy}.}.

Considering that the metric embedded in the brane can be written as $g_{\mu\nu}=\eta_{\mu\nu}+f_{\mu\nu}$ where $\vert\vert f_{\mu\nu}\vert\vert\ll1$, $\eta_{\mu\nu}=diag(-1,1,1,1)$ and assuming an appropriate gauge ($\partial_{\sigma}f^{\sigma}_{\mu}=\frac{1}{2}\partial_{\mu}f$), Eq. \eqref{Eins} can be written as
\begin{equation}
(\nabla^{2}-\partial_{t}^{2})\bar{f}_{\mu\nu}=-\kappa^{2}_{(4)}T_{\mu\nu}^{(0)}-2\kappa^{4}_{(5)}\Pi_{\mu\nu}^{(0)}, \label{Dlambert}
\end{equation}
where $\bar{f}_{,\nu}^{\mu\nu}=0$, $\bar{f}_{\mu\nu}=f_{\mu\nu}-\frac{1}{2}\eta_{\mu\nu}$ and $T_{\mu\nu}^{(0)}$, $\Pi_{\mu\nu}^{(0)}$ are both calculated to zero order in $f_{\mu\nu}$. Note, that we properly retain only contributions of $\Pi_{\mu\nu}$, to study its effects.

Similarly, the conservation equation is maintained $\nabla^{\nu}T_{\mu\nu}^{(0)}=0$, and as we argued previously, we consider non local contributions $\xi_{\mu\nu}=0$, implying that quadratic corrections are also conserved $\nabla^{\nu}\Pi_{\mu\nu}^{(0)}=0$. Roughly speaking, there is not exchange of energy-momentum between the bulk and the brane \citep{Roy}. Also, we assume no fields in the bulk space-time generating that $F_{\mu\nu}=0$. Then Eq. \eqref{Dlambert} can be written as
\begin{equation}
(\nabla^{2}-\partial_{t}^{2})\bar{f}_{\mu\nu}=-\kappa^{2}_{(4)}T_{\mu\nu}^{(0)}-12\kappa^{2}_{(4)}\lambda^{-1}\Pi_{\mu\nu}^{(0)}, \label{Dlambert1}
\end{equation}
where $\kappa^{2}_{(4)}=\lambda\kappa^{4}_{(5)}/6$. In GR limit (the vacuum case), the brane tension is infinite, $\lambda\to\infty$ generating a plane wave solution written as $\bar{f}_{\mu\nu}=A_{\mu\nu}exp(ik_{\alpha}x^{\alpha})$, where $A_{\mu\nu}$ is a constant, symmetric, rank-2 tensor and $k_{\alpha}$ is a constant four-vector known as the \emph{wave vector}. In general, the Green function $G(x^{\mu}-y^{\mu})$, of the d'Alembertian operator is the solution of the wave equation in presence of a delta  function source. Then it is possible to write
\begin{equation}
\bar{f}_{\mu\nu}(x^{\alpha})=-12\kappa^{2}_{(4)}\lambda^{-1}\int_{(4)} G(x^{\alpha}-y^{\alpha})\Pi_{\mu\nu}^{(0)}(y^{\alpha})d^{4}y.
\end{equation}
After some calculations, we obtain the following expression
\begin{equation}
\bar{f}_{ij}(t,\textbf{x})=\mathcal{O}[GR]+\frac{12G_{_{N}}}{\lambda R}\ddot{M}_{ij}(t_{R}),
\end{equation}
where $\mathcal{O}[GR]$ indicates that the GR terms are omitted for simplicity and recovered later. Here, the dots represents derivatives with respect to time. Under the assumption that the spatial extent of the source is negligible compared to the distance between the source and the observer, automatically it is valid the replacement $\vert\textbf{x}-\textbf{y}\vert=R$, being $t_{R}=t-R$ the retarded time and $M_{ij}$ the quadrupole momentum tensor associated with the quadratic energy momentum tensor originated by the brane-world theory
\begin{equation}
M_{ij}=\int_{(3)} y_{i}y_{j}\Pi_{00}^{(0)}d^{3}y. \label{Mij}
\end{equation}
From here, it is straightforward to find the quadrupole radiation equation in the brane-world dynamics as \citep{Peters}
\begin{equation}
P_{_{Br}}=\mathcal{O}[GR]+\frac{6G_{_{N}}}{5\lambda^{2}}\left(\dddot{M}_{ij}\dddot{M}_{ij}-\frac{1}{3}\dddot{M}_{ii}\dddot{M}_{jj}\right)
+\mathcal{I}, \label{redform}
\end{equation}
where Eq. \eqref{redform} shows the projection on its traceless component and $\mathcal{I}$ corresponds to the interference term between $T_{\mu\nu}^{(0)}$ and $\Pi_{\mu\nu}^{(0)}$. 

In the following subsection, we explore the radiation Eq. \eqref{redform} with a binary system \citep{weisberg} with the aim of understand the contributions of the brane-world theory in the gravitational waves emitted by astrophysical objects.

\subsection{Modified power radiation using a binary system.}

Due to the mathematical complexity of Eq. \eqref{Mij} caused by the Dirac deltas, it is most appropriate use the discrete equation of \eqref{Mij}, written as
\begin{equation}
M_{ij}=\sum_{\alpha}m_{\alpha}^{2}y_{\alpha i}y_{\alpha j},
\end{equation}
together with the explicit path of both stars \citep{Peters}
\begin{eqnarray}
\label{}
y^{i}_{1}=d(\phi)\frac{\mu}{m_{1}}\left(cos(\phi),sin(\phi),0\right), \\
\label{}
y^{i}_{2}=d(\phi)\frac{\mu}{m_{2}}\left(-cos(\phi),-sin(\phi),0\right),
\end{eqnarray}
where $\mu$ is the reduced mass defined as $\mu=m_{1}m_{2}/(m_{1}+m_{2})$, being $m_{1}$ and $m_{2}$ the respective mass of the binary stars system. Notice, that the space time structure and in consequence the paths, is only affected by the geometric terms like $\xi_{\mu\nu}$. However, in this case these terms does not play a role.

Now, it is possible to obtain the total average power radiated (recovering GR contribution) over one period of elliptical motion as
\begin{equation}
\left\langle P_{_{Tot}}\right\rangle=\left[(m_{1}+m_{2})+\frac{4\mu m_{1}m_{2}}{\lambda_{_{V}}^{2}}\right]\chi, \label{tot}
\end{equation}
where
\begin{equation}
\chi\equiv\frac{32G^{4}_{_{N}}m_{1}^{2}m_{2}^{2}}{5a^{5}(1
-e^{2})^{7/2}}\left(1+\frac{73}{24}e^{2}+\frac{37}{96}e^{4}\right).
\end{equation}
Notice that $\lambda_{_{V}}$ is the energy value of the brane tension in a defined region, $a$ the semi major axis and $e$ the eccentricity of the ellipse binary system. Note that as a toy model, the interference term will not be studied in this paper. This last expression differs from GR \citep{Peters} only in the second term of the clasp. 

Not less important is to show the equation for radiated power at high energies where brane tension is dominant over the mass distribution, resulting in
\begin{equation}
\left\langle P_{_{Tot}}\right\rangle_{High}=\frac{128G^{4}_{_{N}}\mu m_{1}^{3}m_{2}^{3}}{5a^{5}(1
-e^{2})^{7/2}\lambda_{_{V}}^{2}}\left(1+\frac{73}{24}e^{2}+\frac{37}{96}e^{4}\right).
\end{equation}
Now, from Eq. \eqref{tot} is easily to establish the following bound for the brane tension energy 
\begin{equation}
\lambda_{_{V}}\gg\frac{2m_{1}m_{2}}{m_{1}+m_{2}}=2\mu,
\end{equation}
where the result is only mass depending. With the aim of testing the last result, we probe it using the pulsar binary system PSRB$1913+16$, that is one of the most studied and better known \citep{Peters,1982ApJ,weisberg}. 

Considering the mass of both stars as: $m_{1}=1$.$4414\pm0$.$0002M_{\odot}$ and $m_{2}=1$.$3897\pm0$.$0002M_{\odot}$, it is possible to obtain a numerical upper energy limit for the brane tension as $\lambda_{_{V}}\gg1$.$4150\pm0$.$0002M_{\odot}\sim1$.$5831\times10^{48}EeV$. This result remarks a lower energy lower bound of $\lambda_{_{V}}$, in function of the reduced mass of the astrophysical object. 

\begin{figure}
  \begin{center}
    \includegraphics[width=95mm]{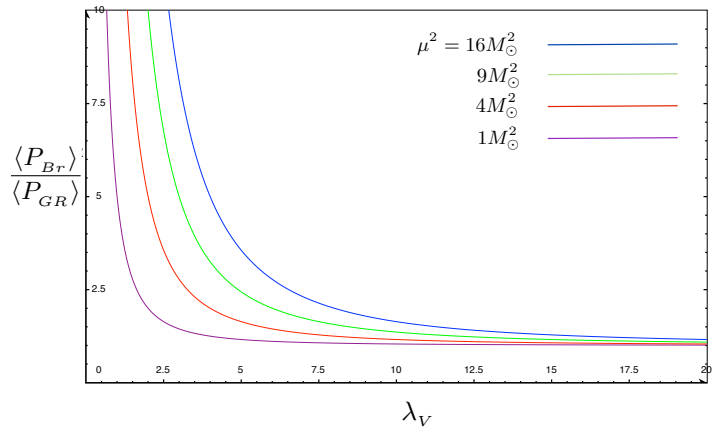}
  \end{center}
  \caption{Behavior of Eq. \eqref{P} for different values of the reduced mass $\mu$. We observe a constant tendency when $\mu/\lambda_{_{V}}\ll1$. On the other hand, when branes dominates the scene, we have the equation $\left\langle P_{_{Br}}\right\rangle=4\mu^{2}\left\langle P_{_{GR}}\right\rangle/\lambda_{_{V}}^{2}$.}
  \label{fig0.0}
\end{figure}

Similarly, notice that
\begin{equation}
\frac{\left\langle P_{_{Br}}\right\rangle}{\left\langle P_{_{GR}}\right\rangle}=1+\frac{4\mu^{2}}{\lambda^{2}_{_{V}}}\simeq1+
1.996\left(\frac{M_{\odot}}{\lambda_{_{V}}}\right)^{2}, \label{P}
\end{equation}
for the binary system PSRB$1913+16$. The behavior is shown in Fig. \ref{fig0.0} for different values of the reduced mass $\mu$. The usual GR result is recovered in the limit

\begin{equation}
  \label{eq1}
   \lim_{\mu/\lambda_{_{V}} \to 0} \left[ 1+\frac{4\mu^{2}}{\lambda_{_{V}}^{2}} \right] = 1 \, ,
\end{equation}
whereas in the opposite case we get
\begin{equation}
  \label{eq1}
   \lim_{\mu/\lambda_{_{V}} \to \infty} \left[ 1+\frac{4\mu^{2}}{\lambda_{_{V}}^{2}} \right] = \frac{4\mu^{2}}{\lambda^{2}_{_{V}}} \, .
\end{equation}

Other interesting exercise is use the data obtained with the system PSRB$1913+16$ and apply to the period equation for gravitational waves in the brane regime (In this case, we extend the results shown in \citep{Peters,1982ApJ}). This equation in high energy regime (where brane effects dominates), can be written as
\begin{equation}
\frac{d\tau_{_{High}}}{dt}=-\frac{384\pi G^{5/3}_{_{N}}m_{1}^{3}m_{2}^{3}}{5\lambda_{_{V}}^{2}}(m_{1}+m_{2})^{-7/3}\left(\frac{\tau}{2\pi}\right)^{-5/3}\mathcal{M},
\end{equation} 
where
\begin{equation}
\mathcal{M}\equiv\left(1+\frac{73}{24}e^{2}+\frac{37}{96}e^{4}\right)(1-e^{2})^{-7/2},
\end{equation}
here $\tau$ is the binary system period. Using the data reported in \cite{1982ApJ}, we obtain the following constraint for brane tension energy: $\lambda_{_{V}}=\sqrt{2}\mu\simeq1\pm0$.$0001M_{\odot}\sim1$.$1194\times10^{48}EeV$, which is in good agreement with the results obtained in the radiated power.   

\section{Discussion and Conclusions.} \label{CR}

Due to the lack of results at microscopic scales in the LHC, the community has decided to look for evidence of brane-worlds in other test. For this reason, in this paper is proposed new ways to detect the existence of branes in different astrophysical tests, or in its case, constraining the brane tension value with the aim of known the validity region of the theory.

Focusing in the gravitational wave case, we explore the machinery behind the modified Einstein's equation and demonstrate the total (GR$+$branes) average power radiated, using the well studied binary system PSRB$1913+16$ as an example. We observe how the new brane term provides with extra amount of average power radiation, with the characteristic that it is mediated with the brane tension. Without loss of generality and assuming that it is valid the same scenario at high energies, it is shown the average power radiated in extreme conditions (where lambda terms is dominant) and compared with the usual case. It is important to remember that in this case, we neglect the non local terms \emph{i.e.} the gravitons that escape from the brane, which generate a reduction in the average power radiated by the source and clearly, an essential changes in the average power radiation equation. 

Also it is shown the modified period by the presence of branes in the high energy regime. In the same way and with the data provided by the binary system, we propose a new lower energy bound for brane tension.
Remarkably, both bounds for the brane tension are compatible with the data provided by nucleosynthesis and other cosmological tests. Even more, the data obtained for the brane tension is compatible with astrophysical test, particularly in stellar stability \cite{Germani}.
 
As a final remark, we think that this attempt can generate a more suitable bound of the brane tension with the capability of being detected in future experiments, principally (under the current experimental scenario) in the evolution of the dynamics of binary systems and other more complex systems composed of neutron stars \cite{Ransom:2014xla}.

\section*{Acknowledgments}

We would like to thank referee for his/her comments. Also, the author was indebted with Luis Ure\~na, Leonardo Ort\'iz, Juan Maga\~na, Yoelsy Leyva, Juan Barranco, Arturo Avelino  and Francisco Linares for reading the first results and for the observations proposed for improving the paper. This work is supported by a CONACyT postdoctoral grant. Instituto Avanzado de Cosmolog\'ia (IAC) and Beyond Standard Theory Group (BeST) collaboration. This work was partially supported by SNI-CONACyT (M\'exico).

\bibliographystyle{apsrev}

\bibliography{librero1}

\end{document}